\documentclass[aps,prl,twocolumn,footinbib,superscriptaddress]{revtex4}
\usepackage{graphicx}%
\usepackage{siunitx}
\usepackage{bm}
\usepackage[caption=false]{subfig}
\usepackage{mathtools}
\usepackage{multirow}
\usepackage{amssymb}
\usepackage{braket}
\usepackage{amsmath}
\usepackage{placeins}
\usepackage{cancel}
\usepackage{xcolor}
\usepackage{spreadtab}
\usepackage[colorlinks=true,
            linkcolor=blue,
	        citecolor=blue,
	         urlcolor=blue]{hyperref}

\begin{document}

\title{Accurate prediction of Hall mobilities in two-dimensional materials through gauge-covariant quadrupolar contributions}

\author{Samuel Ponc\'e}
\email{samuel.ponce@uclouvain.be}
\affiliation{%
Institute of Condensed Matter and Nanosciences (IMCN), Universit\'e catholique de Louvain, Chemin des \'Etoiles 8, B-1348 Louvain-la-Neuve, Belgium
}%
\affiliation{%
Theory and Simulation of Materials (THEOS), \'Ecole Polytechnique F\'ed\'erale de Lausanne,
CH-1015 Lausanne, Switzerland
}%
\author{Miquel Royo}
\affiliation{%
Institut de Ci\`encia de Materials de Barcelona (ICMAB-CSIC), Campus UAB, 08193 Bellaterra, Spain
}%
\author{Marco Gibertini}
\affiliation{%
Dipartimento di Scienze Fisiche, Informatiche e Matematiche, Universit\`a di Modena e Reggio Emilia, Via Campi 213/a I-41125 Modena, Italy
}%
\affiliation{%
Centro S3, Istituto Nanoscienze-CNR, Via Campi 213/a, I-41125 Modena, Italy
}%
\author{Nicola Marzari}
\affiliation{%
Theory and Simulation of Materials (THEOS), \'Ecole Polytechnique F\'ed\'erale de Lausanne,
CH-1015 Lausanne, Switzerland
}%
\affiliation{%
National Centre for Computational Design and Discovery of Novel Materials (MARVEL), \'Ecole Polytechnique F\'ed\'erale de Lausanne, CH-1015 Lausanne, Switzerland
}%
\author{Massimiliano Stengel}
\affiliation{%
Institut de Ci\`encia de Materials de Barcelona (ICMAB-CSIC), Campus UAB, 08193 Bellaterra, Spain
}%
\affiliation{Instituci\'o Catalana de Recerca i Estudis Avançats (ICREA), Pg. Llu\'is Companys, 23, 08010 Barcelona, Spain}

\date{\today}

\begin{abstract}
Despite considerable efforts, accurate computations of electron-phonon and carrier transport properties of low-dimensional materials from first principles have remained elusive.
By building on recent advances in the description of long-range electrostatics, we develop a general approach to the calculation of electron-phonon couplings in two-dimensional materials.
We show that the nonanalytic behavior of the electron-phonon matrix elements depends on the Wannier gauge, but that a missing Berry connection restores invariance
to quadrupolar order.
We showcase these contributions in a MoS$_2$ monolayer, calculating intrinsic drift and Hall mobilities with precise Wannier interpolations. 
We also find that the contributions of dynamical quadrupoles to the scattering potential are essential, and that their neglect leads to errors of 23\% and 76\% in the room temperature electron and hole Hall mobilities, respectively.
\end{abstract}

\maketitle

Charge transport in two-dimensional (2D) materials lies at the heart of many technological applications
ranging from transistors~\cite{Radisavljevic2011}, solar cells~\cite{Pospischil2014,Xia2021a}, emitters~\cite{Ren2019}, and photodetectors~\cite{Jiang2021}.
It is therefore desirable to understand the different scattering mechanisms that govern carriers transport.
In the case of high-quality samples with low doping concentrations, electron-phonon scattering is the dominant mechanism limiting carrier mobilities~\cite{Ponce2020}.
Given the current challenges with the experimental probing of electron-phonon 
interactions in 2D, theoretical studies based on \textit{ab-initio} simulations are crucial for future progress.

Recent advances in first-principles calculations of mobilities~\cite{Ponce2020} have made it possible to study accurately and without empirical parameters electronic transport in semiconductors.
At the core of such calculations are the electron-phonon matrix elements $g_{mn\nu}(\mathbf{k},\mathbf{q})$, that describe the scattering amplitudes from an initial electron Bloch state $n\mathbf{k}$ to a final state $m\mathbf{k+q}$ via the emission or absorption of a phonon with frequency $\omega_{\mathbf{q}\nu}$.
One of the major challenges lies in achieving numerically converged results, which require sampling of $g_{mn\nu}(\mathbf{k},\mathbf{q})$ on ultra-dense momentum grids~\cite{Ponce2018}.
To make the problem tractable, an accurate way forward consists in explicitly computing the electron-phonon matrix element 
on coarse grids via state-of-the-art density-functional perturbation theory (DFPT)~\cite{Gonze1997a,Baroni2001} methods, followed by Fourier interpolation to these ultra-dense grids~\cite{Giustino2007,Eiguren2008,Calandra2010,Brunin2020}.

However, this approach requires further care when dealing with the long-range electrostatic fields that arise near the Brillouin-zone center in semiconductors and insulators.
The leading Fr\"ohlich interaction~\cite{Vogl1976}, together with higher-order multipolar contributions~\cite{Brunin2020,Brunin2020a,Jhalani2020,Park2020a,Ponce2021}, make the electron-phonon interactions nonanalytic in the long-wavelength limit, which precludes straightforward application of Fourier interpolations~\cite{Verdi2015,Sjakste2015}.
The strategy to tackle such problem rests on a formal analysis of the nonanalytic properties of the scattering potential~\cite{Brunin2020,Brunin2020a} and is now well established in the context of 3D crystals~\cite{Giustino2017}.
By contrast, the long-range electrostatic problem in 2D materials has not been thoroughly investigated.
A major advance in this direction came with a formulation of the long-range interactions that accounts for the effect of the in-plane dipoles~\cite{Sohier2016,Sohier2017} and relies on out-of-plane Coulomb truncations~\cite{Sohier2017a}.
Nonetheless, the treatment of such interactions has been based on dielectric analogues, which correctly capture the leading Fr\"ohlich-like term, but miss higher multipoles, that are instead fully included in the DFPT framework~\cite{Sohier2017}. 
Efforts in this direction have been reported recently~\cite{Deng2021,Zhang2022}, but a fundamental understanding of higher-order multipolar couplings in 2D is still missing.

Treating the higher-order electrostatics accurately is enough for methods that are based on the interpolation of the scattering potential $V_{\mathbf{q}\nu}$~\cite{Brunin2020a}.
Their main drawback lies in that a massive amount of plane-wave components and electronic eigenstates need to 
be computed, stored and processed on a dense grid of \textbf{k}-points spanning the Brillouin zone. 
To avoid this, methods where the $g_{mn\nu}(\mathbf{k},\mathbf{q})$ rather than $V_{\mathbf{q}\nu}$ are interpolated are often preferred.
The key issue is that the electron-phonon matrix elements are not smooth in \textbf{q}, as they depend on the random gauge of the electronic orbitals. 
Modern Wannier-function techniques~\cite{Marzari2012} become then key to ensuring optimal smoothness across the Brillouin zone and enable efficient interpolations. 
However the choice of Wannier functions is nonunique and this can lead to an undesirable arbitrariness in the quality of the interpolation, which depends on the specific conventions used in the numerical implementation in an uncontrolled way.
Earlier work has dealt almost exclusively on the leading-order Fr\"ohlich Hamiltonian where this issue does not apply~\cite{Verdi2015,Sjakste2015}.
Yet, given the qualitative importance of higher-order corrections~\cite{Brunin2020a,Jhalani2020,Park2020a,Ponce2021}, understanding the role of the Wannier gauge
is mandatory to ensure consistency of the method.

Here, we find that the Wannier gauge affects the long-wavelength behavior of the $g_{mn\nu}(\mathbf{k},\mathbf{q})$ in a nontrivial way, and that contrary to widespread assumptions, the long-range matrix elements are not gauge-covariant. 
Nonanalyticities due to the \textbf{k}-dispersion of the Wannier gauge appear at the same order as the quadrupolar contributions to the scattering potential, and are therefore essential to incorporate in any beyond-Fr\"ohlich Wannier-based approach.
Importantly, we show that these terms acquire the form of a Berry connection (the position operator represented on a Wannier basis), and demonstrate its importance at leading order in \textbf{q}.
We focus here on drift and Hall carrier mobilities, and take monolayer MoS$_2$ as a paradigmatic case study; 
such choice is motivated by its technological relevance, and the availability of many theoretical and experimental data for this system. 
Results for five additional representative 2D crystals and a 3D material are reported in our accompanying paper~\cite{Ponce2022}, where we identify a non-trivial temperature evolution of the Hall hole mobility in InSe.
In particular, we show that mobilities are strongly affected by long-range effects beyond Fr\"ohlich, and that a correct treatment leads to a room-temperature electron mobility 25\% smaller than previous reports~\cite{Deng2021}.
This highlights the importance of including this correct quadrupolar coupling in the description of electron-phonon interactions in semiconducting low-dimensional materials.

\begin{figure}[t]
  \centering
  \includegraphics[width=0.99\linewidth]{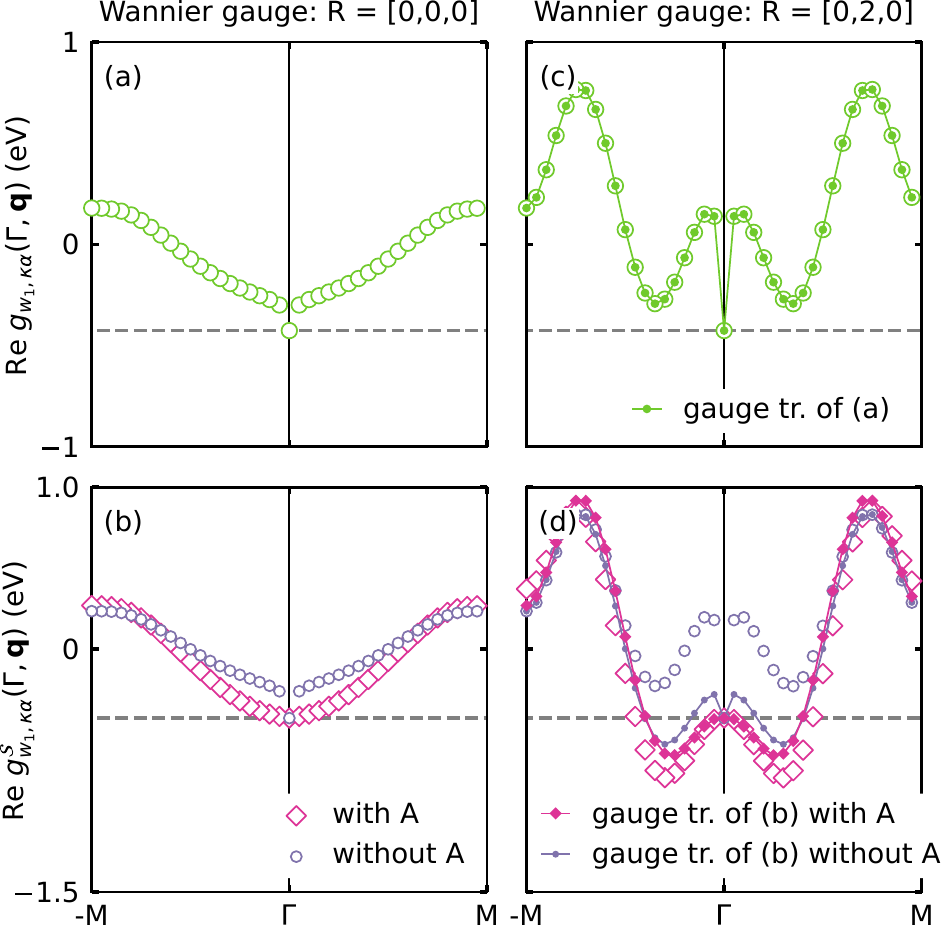}
  \caption{\label{fig1}
Real part of the electron-phonon matrix element $g$ as a function of phonon momentum $\mathbf{q}$ of MoS$_2$ monolayer for the first Wannier function ($w_1$) describing the maximum of the valence band at the $\mathbf{k}$=$\Gamma$ point and for the displacement of the Mo atom in the direction of the $M$ point.
(a) Direct DFPT results in the smooth Wannier gauge where the Wannier centers are located in the primitive cell as well as the short range component (b) with and without the new gauge restoring term $A$.
Note that at \textbf{q}=\textbf{0}, the DFPT calculations give access only to the short-range part and differ from the $\mathbf{q}\to\mathbf{0}$ limit. 
Only with the new $A$ term is the short range part continuous and analytic in the long-wavelength limit. 
(c,d) Same results as (a,b) but where the Wannier centers are located two lattice vectors away in the $M$ direction.
The lines with filled symbols are obtained by multiplying the data in (a) and (b) with the gauge transformation associated with the \textbf{R}=[0,2,0] translation. 
The direct calculation with the $A$ term gives results which fulfill gauge covariance close to the zone center, in contrast with the result without the A term.  
The local fields response to an electric field $V^{\mathcal{E}}$ is included but has a negligible effect.      
}
\end{figure}

To deal with higher order terms, we decompose $g_{mn\nu}(\mathbf{k},\mathbf{q})$ into short- ($\mathcal{S}$) and long-range ($\mathcal{L}$) contributions, where the long-range part can be formally expressed in terms of a scattering potential $V_{\mathbf{q}\kappa\alpha}^{\mathcal{L}}$, referring to the displacement of atom $\kappa$ in the Cartesian direction $\alpha$, as:
\begin{multline}\label{eq:multipole}
g^\mathcal{L}_{mn\nu}(\mathbf{k,q}) = \Big[ \frac{\hbar}{2 \omega_{\nu}(\mathbf{q})}\Big]^{\frac{1}{2}}  \sum_{\kappa\alpha}  \frac{e_{\kappa\alpha\nu}(\mathbf{q})}{\sqrt{M_{\kappa}}} \sum_{sp}\\
\times  U_{ms\mathbf{k+q}}   \langle u_{s\mathbf{k+q}}^{\rm W}| V_{\mathbf{q}\kappa\alpha}^{\mathcal{L}} | u_{p\mathbf{k}}^{\rm W} \rangle U_{pn\mathbf{k}}^{\dagger}.
\end{multline}
Here, we have changed representation to the maximally localized Wannier gauge~\cite{Marzari2012} to guarantee a smooth behavior in $\mathbf{q}$ of the cell-periodic part of the Bloch eigenstates $\ket{u_{n\mathbf{k}}} = \sum_p U_{np\mathbf{k}}^{*} \ket{u^{\rm W}_{p\mathbf{k}}}$, where $e_{\kappa\alpha\nu}$ are the dynamical matrix eigenstates corresponding to a phonon mode $\nu$ of momentum $\mathbf{q}$.

The mirror-even part of the long-range scattering potential in a quasi-2D crystal can be written~\cite{Ponce2022} as
\begin{equation}\label{eq:dipole}
V_{\mathbf{q}\kappa\alpha}^{\mathcal{L}}(\mathbf{r}) =  \frac{e}{S} \frac{2 \pi f(q)}{q} \bigg[ \frac{i\mathbf{q} \cdot \boldsymbol{\mathcal{Z}}_{\kappa\alpha}^{\parallel}(\mathbf{q})}{\epsilon^{\parallel}(\mathbf{q})} \tilde{\varphi}_{\mathbf{q}}^\parallel(\mathbf{r}) \bigg] 
e^{-i \mathbf{q} \cdot \boldsymbol{\tau}_\kappa},
\end{equation}
where $S$ is the unit-cell area, $\boldsymbol{\tau}_{\kappa}$ stands for the position of atom $\kappa$ within the cell, and $q=|\mathbf{q}|$.
We neglect the mirror-odd contribution to $V_{\mathbf{q}\kappa\alpha}^{\mathcal{L}}(\mathbf{r})$ since it does not contribute to the mobility of the system considered here~\cite{Ponce2022}.
The range separation function $f(q) = 1 - \tanh(q L / 2)$ is a low-pass Fourier filter that ensures the macroscopic character of the potential, where the parameter $L$ defines the length scale~\cite{Royo2021}.
Note that the choice of $L$ is, to a large extent, arbitrary and can be tuned to maximize the numerical efficiency of the interpolation~\footnote{Here we follow the prescriptions of Ref.~\onlinecite{Royo2021}, based on the analysis of the real-space interatomic force constants; for MoS$_2$, we find that the optimal range separation parameter is $L = 10.5$~bohr.}.
Equation~\eqref{eq:dipole} can be intuitively interpreted as the bare long-range Coulomb kernel in 2D, $\nu^{\textrm{2D}}(q) = 2 \pi f(q)/q$, multiplied by a screened surface polarization charge and by a form factor $\tilde{\varphi}_{\mathbf{q}}^\parallel(\mathbf{r})$. 
The latter reflects the fact that the electrostatic potentials produced by a modulated plane of charge are nonuniform along the out-of-plane direction~\cite{Royo2021};
note that the form factor depends on $\mathbf{r}$ and not only on the out-of-plane distance $z$ due to local fields (see Eq.~\eqref{eq:uexp} below).
In turn, the screened charge is written as the divergence of the polarization field associated with the displacement 
of atom $\kappa$ along the in-plane Cartesian direction $\alpha$, $\boldsymbol{\mathcal{Z}}_{\kappa\alpha}^{\parallel}(\mathbf{q})$, 
divided by the macroscopic in-plane dielectric function $\epsilon^{\parallel}(\mathbf{q}) = 1 + \nu^{\textrm{2D}}(q) \alpha^{\parallel}(\mathbf{q})$, 
where $\alpha^{\parallel}(\mathbf{q})$ is the macroscopic in-plane polarizability~\footnote{The macroscopic 
in-plane polarizability can be obtained from the in-plane dielectric constant $\varepsilon_{\alpha\beta}$ of 
an artificially periodic stack of monolayers with spacing $c$ through $\alpha^{\parallel}(\mathbf{q}) = (c/4\pi)\sum_{\alpha\beta}q_\alpha(\varepsilon_{\alpha\beta} - \delta_{\alpha\beta}) q_\beta$.}.
In the long-wavelength limit, $\boldsymbol{\mathcal{Z}}_{\kappa\alpha}^{\parallel}(\mathbf{q})$ 
can be conveniently expressed in a multipole expansion as  $\boldsymbol{\mathcal{Z}}_{\kappa\alpha}^{\parallel}(\mathbf{q}) \equiv Z_{\kappa\alpha\beta} - i \sum_{\gamma}\frac{q_\gamma}{2}(Q_{\kappa\alpha\beta\gamma} \!-\! \delta_{\beta\gamma} Q_{\kappa\alpha zz})+\mathcal{O}(\mathbf{q}^2)$.
$Z_{\kappa\alpha\beta}$ are the dynamical Born effective charge tensors~\cite{Gonze1997a,Baroni2001}, and $Q_{\kappa\alpha\beta\gamma}$ the dynamical quadrupole tensors~\cite{Stengel2013,Royo2019}.

\begin{figure}[t]
  \centering
  \includegraphics[width=0.9\linewidth]{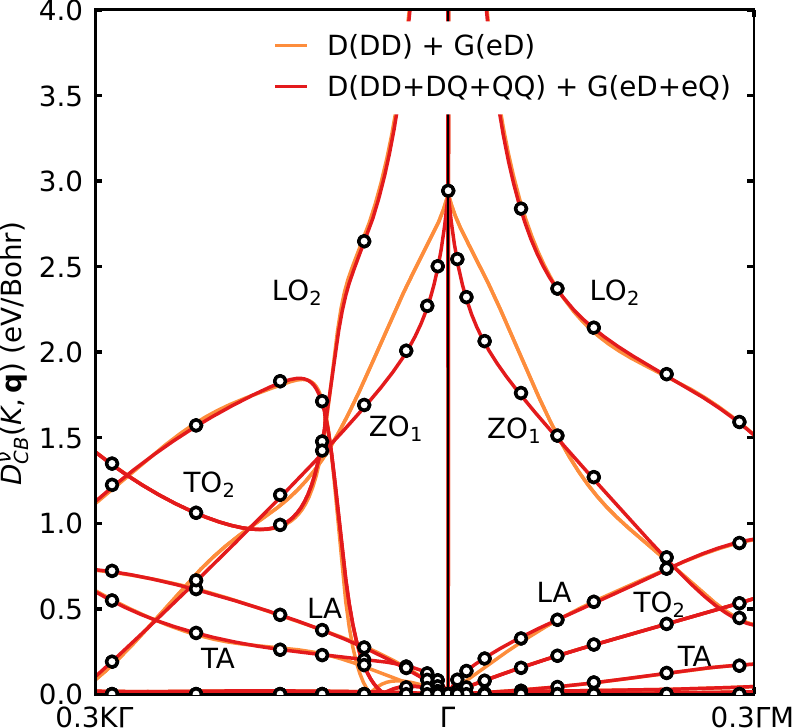}
  \caption{\label{fig2}
Deformation potential of MoS$_2$ monolayer at the conduction band at \textbf{k}=\textbf{K} calculated with DFPT in the correct electrostatic open-boundary conditions~\cite{Sohier2017} (black dots) compared with Fourier interpolation where the long-range part of the electron-phonon matrix elements (G) includes monopole-dipole (eD) and monopole-quadrupole (eQ) and the dynamical matrix (D) includes dipole-dipole (DD), dipole-quadrupole (DQ) and quadrupole-quadrupole (QQ) contributions.
}
\end{figure}

The explicit expression in Eq.~\eqref{eq:dipole} for the scattering potential allows to construct the corresponding matrix element in Eq.~\eqref{eq:multipole}, which can be subtracted from the \textit{ab-initio} results, making these short-ranged and analytic in $\mathbf{q}$, and 
amenable to accurate Fourier-Wannier interpolations.
Adding the non-analytic expressions back allows then to obtain the electron-phonon matrix elements over arbitrarily dense grids~\cite{Verdi2015,Sjakste2015}.
Importantly, in order to recover the correct $\mathbf{q} \rightarrow 0$ expansion of Eq.~\eqref{eq:multipole}, one also needs the expansion to first order of the matrix elements of the form factors,
\begin{equation}\label{eq:uexp}
\langle u_{s\mathbf{k+q}}^{\rm W}| \tilde{\varphi}_{\mathbf{q}}^\parallel | u_{p\mathbf{k}}^{\rm W} \rangle \approx \delta_{sp} \!+\! i \mathbf{q} \! \cdot \!
  \Big[  \langle u_{s\mathbf{k}}^{\rm W}| V^{\boldsymbol{\mathcal{E}}^\parallel}| u_{p\mathbf{k}}^{\rm W} \rangle  \!+\!  \mathbf{A}_{sp\mathbf{k}}^{\text{W}} \Big],
\end{equation}
where we have exploited that the Wannier gauge is smooth everywhere in the Brillouin zone.
The expansion in Eq.~\eqref{eq:uexp} therefore involves two terms beyond Ref.~\cite{Verdi2015}.
The first one involves the matrix elements of the self-consistent potential (``local fields'') in response to a uniform electric field $\boldsymbol{\mathcal{E}}$ that is oriented parallel to the layer plane. 
A similar contribution has already been identified in the 3D case~\cite{Brunin2020a,Brunin2020} and found to be small;
 we find that the impact of local fields is negligible also in 2D materials, with an effect below 0.1\% on the mobility.

The second contribution consists in a novel and previously unreported term arising from the Berry connection $\mathbf{A}_{sp\mathbf{k}}^{\text{W}}$.
This term is crucial as it ensures the smoothness of the short-range electron-phonon matrix elements in the Wannier gauge, see Fig.~\ref{fig1}(b).
Additionally, this term restores covariance to lowest order in \textbf{q} under Wannier gauge transformations.
This is shown in Fig.~\ref{fig1}(d) by considering a different Wannier gauge obtained by displacing all Wannier centers by a lattice vector \textbf{R},
which introduces a phase factor to the Wannier rotation matrix $U_{sp\mathbf{k}} e^{- i \mathbf{k} \cdot \mathbf{R}}$, shown as a line with filled symbols. 
Once the Berry connection term is applied, the results match exactly the gauge transformed ones in the $\mathbf{q}\to\mathbf{0}$ limit;  
  without it, there is a significant amplification of the zone-center discontinuity and a different long-wavelength limit.

\begin{figure}[t]
  \centering
  \includegraphics[width=0.95\linewidth]{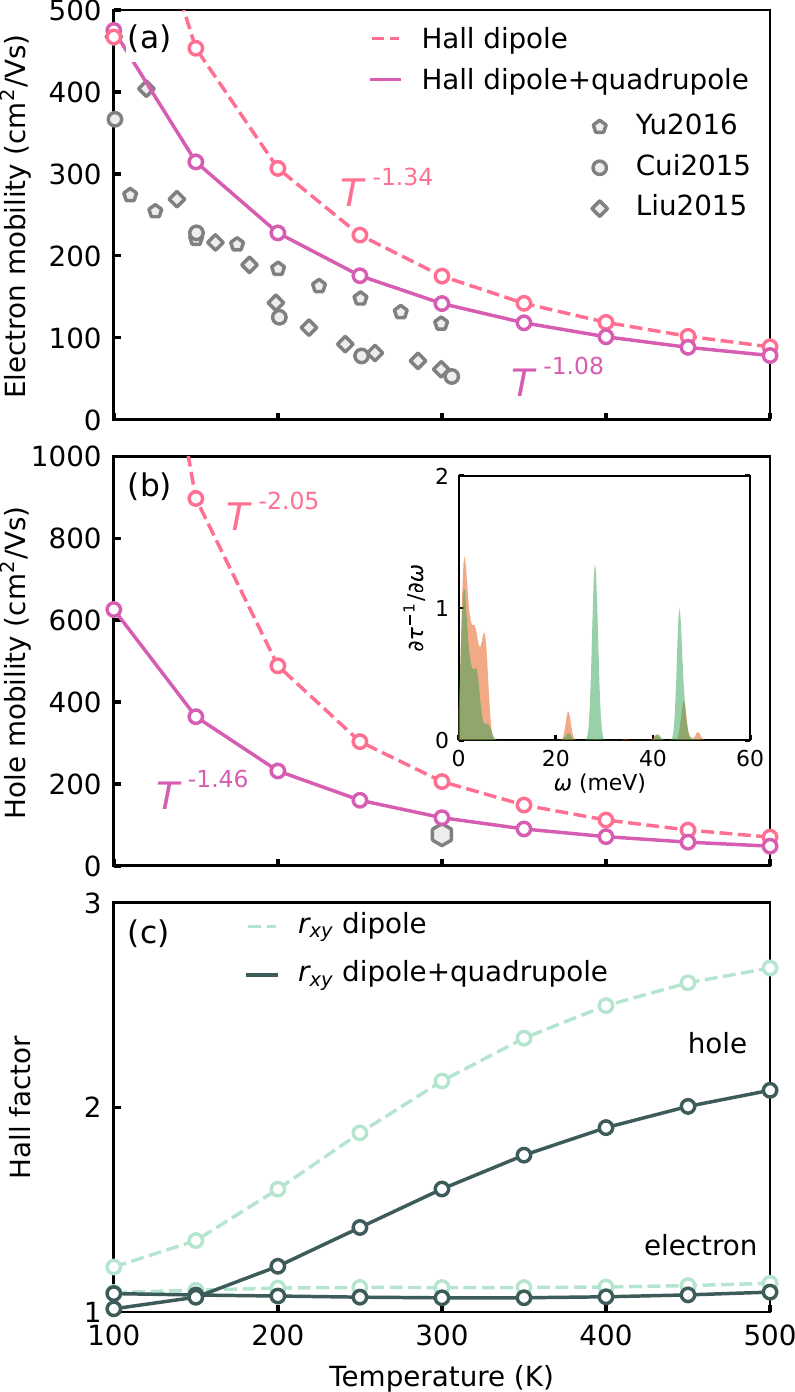}
  \caption{\label{fig3}
Temperature dependence of the Hall carrier mobility for (a) electrons and (b) holes in MoS$_2$ as well as (c) the electron and hole Hall factor.
The dashed  (solid) lines represent the mobilities calculated using electron-phonon interactions considering only the dipole or dipole and quadrupole contributions; the temperature exponent for the mobility is also reported.
The gray symbols refer to experimental data from Refs.~\cite{Yu2016,Cui2015,Liu2015c,Momose2018}.
The inset in (b) provides the spectral decomposition of the electron (orange) and hole (green) scattering rates at 300~K as a function of phonon energy, and calculated as angular averages for carriers at an energy $3/2 k_BT$ = 39~meV from the band edge.
}
\end{figure}

With these capabilities we are now in a position to obtain the low-field phonon-limited carrier mobility in the presence of a vanishing magnetic field \textbf{B} by solving the Boltzmann transport equation (BTE)~\cite{Ponce2020,Macheda2020,Ponce2021}.
From this solution we can also compute the Hall factor $r_{\alpha\beta}(\hat{\mathbf{B}})$ as the ratio between mobilities with/without magnetic field~\cite{Reggiani1983,Popovic1991,Macheda2018}.
We perform our calculations using the \textsc{EPW}~\cite{Giustino2007,Ponce2016a}, \textsc{Wannier90}~\cite{Pizzi2020}, and \textsc{Quantum ESPRESSO}~\cite{Giannozzi2017} packages, including spin-orbit coupling (SOC) and using fully converged computational parameters~\footnote{
We use fully relativistic norm-conserving Perdew-Burke-Ernzerhof (PBE) pseudopotentials~\cite{Perdew1996,Hamann2013,Setten2018} which include the $4s^2$, $4p^6$, $4d^5$, $5s^1$ as valence states for Mo and $3s^2$, $3p^4$ as valence states for S.
The electron wave functions are expanded in a plane-wave basis set with kinetic energy
cutoff of 140~Ry, and the Brillouin zone is sampled using a homogeneous $\Gamma$-centred 18$\times$18$\times$1 mesh.}.
In addition, we use the long-wave driver of \textsc{Abinit}~\cite{Gonze2016,Gonze2020} to compute the quadrupoles~\cite{Royo2019} in absence of SOC. 
The linearly-independent quadrupoles obtained are Q$_{\kappa yyy}$ = 5.533~$e\cdot$bohr for Mo, and
Q$_{\kappa y yy} =  0.391$, Q$_{\kappa yyz} = -0.174$,
Q$_{\kappa zxx}$ = 7.858, and Q$_{\kappa zzz}$ = 0.230~$e\cdot$bohr for one of the two S atoms.

In MoS$_2$, there are three sets of mirror-even branches that contribute crucially to the mobilities: the zone-center infrared and Raman active E$^{'}$ mode associated with an in-plane out-of-phase movement of the Mo and S atoms, which splits into LO$_2$ and TO$_2$ branches at finite momentum; the Raman active A$_1^{'}$ mode associated with the out-of-plane motion of the sulfur atoms while the molybdenum atoms remain fixed (ZO$_1$ branch); and the zone-center E$^{'}$ acoustic mode that splits into the LA and TA branches at finite momentum.
In Fig.~\ref{fig2} we show the Wannier-interpolated deformation potentials~\cite{Zollner1990,Sjakste2015} for all the phonon branches, and compare these with direct DFPT calculations (black circles).
The mirror-odd ZA, LO$_1$, TO$_1$, and ZO$_2$ modes are inactive.
We show in Fig.~\ref{fig2} that the addition of dynamical quadrupoles is essential to recover the correct ZO$_1$ deformation potential.
In Ref.~\onlinecite{Ponce2022}, we also show how the dipole approximation yields deformation potentials with quantitatively and qualitatively incorrect long-wavelength dispersions for different phonon branches for each type of 2D material considered.
In all these cases, the inclusion of the quadrupolar fields recovers the correct dispersions.

Finally, we show in Fig.~\ref{fig3} the intrinsic Hall electron and hole mobilities
that are obtained considering only dipoles, or dipoles and quadrupoles, on dense 500$\times$500 \textbf{k} and \textbf{q} momentum grids.
The latter yield room temperature Hall mobilities of 143~cm$^2$/Vs and 113~cm$^2$/Vs for electrons and holes, respectively.
Importantly, the most noticeable result is the impact that the inclusion of the quadrupoles in the interpolation entails on the final results:
the room-temperature Hall mobility is reduced by 23\% for electrons and by 76\% for holes, when compared to the dipoles-only case.
In Table~III of our accompanying manuscript Ref.~\onlinecite{Ponce2022} we compare the effects of including quadrupole corrections in the dynamical matrix or electron-phonon matrix elements for the computed mobilities, and we find that including it only in the electron-phonon matrix elements has the largest impact. 
Finally, we find that omitting the Berry connection term has a small effect on the Hall mobility of MoS$_2$ and yields 142~cm$^2$/Vs and 117~cm$^2$/Vs for electrons and holes, respectively.
In other 2D materials, such as BN or InSe studied in our accompanying paper~\cite{Ponce2022}, the Berry connection plays a crucial role and is therefore required. 
This is true also for 3D bulk materials, with a spectacular impact in the case of SrO~\cite{Ponce2022}.

The comparison with the intrinsic drift mobilities of 132~cm$^2$/Vs and 73~cm$^2$/Vs (not shown in the figure), reveals that the holes are more strongly affected by the presence of a magnetic field.
Besides, in Fig.~\ref{fig3}(c) we show how the hole Hall factor strongly increases with temperature, doubling from 100~K to 500~K, while the electron Hall factor is almost unaffected.
In the inset of Fig.~\ref{fig3}(b) we show that 87\% of the electron mobility of this MoS$_2$ monolayer is limited by acoustic scattering, while 52\% of the hole mobility is due to optical scattering.
Since quadrupoles mostly affect the ZO$_1$ optical mode, this clarifies why the hole mobility is affected more than the electron mobility.

For the room-temperature electron mobility, experimental values range from 23 to 217~cm$^2$/Vs~\cite{Yu2014,Sanne2015,Kang2015,Cui2015,Liu2015c,Yu2016}, and typically rely on the use of high-permittivity gate dielectrics.
The goal of such dielectric is to suppress Coulomb scattering by immersing the MoS$_2$ monolayer in a high-$\kappa$ dielectric environment to boost mobility.
There is a large spread in the reported theoretical literature, with room-temperature electron mobilities in the range 97-410~cm$^2$/Vs~\cite{Kaasbjerg2012a,Li2013,Kaasbjerg2013,Zhang2014,Li2015,Gunst2016,Sohier2018,Guo2019,Gaddemane2019,Deng2021}.
In the case of the hole mobilities instead, the only previous calculation did not consider SOC, yielding a room-temperature value of 26~cm$^2$/Vs~\cite{Guo2019}, underestimating by a factor of three the 76~cm$^2$/Vs experimental value~\cite{Momose2018}, and by a factor of four the present result.

We note that the decrease in hole mobility occurring for temperatures above 200~K in Fig.~\ref{fig3}(c) is due to the $\Gamma$ valley being only 52~meV below the valence-band maximum and therefore becoming thermally populated at higher temperatures. 
In addition to the SOC, and as seen in Fig.~\ref{fig3}, the role of dynamical quadrupoles is crucial and delivers predicted drift and Hall mobility close to experimental values, with also the exponent of the power-law temperature decrease in the electron mobility going from $-1.34$ to $-1.08$, closer to measurements.

In conclusion, we have developed and implemented a conceptual and numerical framework to accurately determine electron-phonon couplings in 2D materials on ultra-dense momentum grids, including full 2D electrostatics, quadrupoles, and SOC, highlighting how those are critical for the electron and hole mobilities of 2D materials like MoS$_2$.
Remarkably, we have pointed out a previously unreported missing Wannier gauge covariance for the interpolation procedure, and showed how to restore consistency by including a Berry connection term. 
Given the major role played by low-dimensional semiconductors in the post-silicon roadmap~\cite{IRDS2001}, we believe that understanding of the physical effects described here will be of great value in the characterization and engineering of these promising and challenging materials. 

\begin{acknowledgments}
S.P. acknowledges support from the F.R.S.-FNRS as well as from the European Unions Horizon 2020 Research and Innovation Program, under the Marie
Skłodowska-Curie Grant Agreement (SELPH2D, No. 839217);
N.M. acknowledges support from the Swiss National Science Foundation and the NCCR MARVEL;
M.G. acknowledges support from the Italian Ministry for University and Research through the Levi-Montalcini program;
M.S. and M.R. acknowledge support from Ministerio de Ciencia y Innovaci\'on (MICINN-Spain) through Grant No. PID2019-108573GB-C22; from Severo Ochoa FUNFUTURE center of excellence (CEX2019-000917-S); from Generalitat de Catalunya (Grant No. 2017 SGR1506); and from the European Research Council (ERC) under the European Union's Horizon 2020 research and innovation program (Grant Agreement No. 724529).
Computational resources have been provided by the PRACE-21 resources MareNostrum at BSC-CNS, the Supercomputing Center of Galicia (CESGA), and by the Consortium des \'Equipements de Calcul Intensif (C\'ECI), funded by the Fonds de la Recherche Scientifique de Belgique (F.R.S.-FNRS) under Grant No. 2.5020.11 and by the Walloon Region
as well as computational resources awarded on the Belgian share of the EuroHPC LUMI supercomputer.
\end{acknowledgments}


\begin{thebibliography}{61}
\expandafter\ifx\csname natexlab\endcsname\relax\def\natexlab#1{#1}\fi
\expandafter\ifx\csname bibnamefont\endcsname\relax
  \def\bibnamefont#1{#1}\fi
\expandafter\ifx\csname bibfnamefont\endcsname\relax
  \def\bibfnamefont#1{#1}\fi
\expandafter\ifx\csname citenamefont\endcsname\relax
  \def\citenamefont#1{#1}\fi
\expandafter\ifx\csname url\endcsname\relax
  \def\url#1{\texttt{#1}}\fi
\expandafter\ifx\csname urlprefix\endcsname\relax\def\urlprefix{URL }\fi
\providecommand{\bibinfo}[2]{#2}
\providecommand{\eprint}[2][]{\url{#2}}

\bibitem[{\citenamefont{Radisavljevic et~al.}(2011)\citenamefont{Radisavljevic,
  Radenovic, Brivio, Giacometti, and Kis}}]{Radisavljevic2011}
\bibinfo{author}{\bibfnamefont{B.}~\bibnamefont{Radisavljevic}},
  \bibinfo{author}{\bibfnamefont{A.}~\bibnamefont{Radenovic}},
  \bibinfo{author}{\bibfnamefont{J.}~\bibnamefont{Brivio}},
  \bibinfo{author}{\bibfnamefont{V.}~\bibnamefont{Giacometti}},
  \bibnamefont{and} \bibinfo{author}{\bibfnamefont{A.}~\bibnamefont{Kis}},
  \bibinfo{journal}{Nat. Nanotechnol.} \textbf{\bibinfo{volume}{6}},
  \bibinfo{pages}{147} (\bibinfo{year}{2011}).

\bibitem[{\citenamefont{Pospischil et~al.}(2014)\citenamefont{Pospischil,
  Furchi, and Mueller}}]{Pospischil2014}
\bibinfo{author}{\bibfnamefont{A.}~\bibnamefont{Pospischil}},
  \bibinfo{author}{\bibfnamefont{M.~M.} \bibnamefont{Furchi}},
  \bibnamefont{and} \bibinfo{author}{\bibfnamefont{T.}~\bibnamefont{Mueller}},
  \bibinfo{journal}{Nature Nanotechnology} \textbf{\bibinfo{volume}{9}}
  (\bibinfo{year}{2014}).

\bibitem[{\citenamefont{Xia et~al.}(2021)\citenamefont{Xia, Peng, Ponc\'e,
  Patel, Wright, Crothers, Rothemann, Borchert, Milot, Kraus
  et~al.}}]{Xia2021a}
\bibinfo{author}{\bibfnamefont{C.}~\bibnamefont{Xia}},
  \bibinfo{author}{\bibfnamefont{J.}~\bibnamefont{Peng}},
  \bibinfo{author}{\bibfnamefont{S.}~\bibnamefont{Ponc\'e}},
  \bibinfo{author}{\bibfnamefont{J.~B.} \bibnamefont{Patel}},
  \bibinfo{author}{\bibfnamefont{A.~D.} \bibnamefont{Wright}},
  \bibinfo{author}{\bibfnamefont{T.~W.} \bibnamefont{Crothers}},
  \bibinfo{author}{\bibfnamefont{M.~U.} \bibnamefont{Rothemann}},
  \bibinfo{author}{\bibfnamefont{J.}~\bibnamefont{Borchert}},
  \bibinfo{author}{\bibfnamefont{R.~L.} \bibnamefont{Milot}},
  \bibinfo{author}{\bibfnamefont{H.}~\bibnamefont{Kraus}},
  \bibnamefont{et~al.}, \bibinfo{journal}{J. Phys. Chem. Lett.}
  \textbf{\bibinfo{volume}{12}}, \bibinfo{pages}{3607} (\bibinfo{year}{2021}).

\bibitem[{\citenamefont{Ren et~al.}(2019)\citenamefont{Ren, Tan, and
  Zhang}}]{Ren2019}
\bibinfo{author}{\bibfnamefont{S.}~\bibnamefont{Ren}},
  \bibinfo{author}{\bibfnamefont{Q.}~\bibnamefont{Tan}}, \bibnamefont{and}
  \bibinfo{author}{\bibfnamefont{J.}~\bibnamefont{Zhang}},
  \bibinfo{journal}{Journal of Semiconductors} \textbf{\bibinfo{volume}{40}},
  \bibinfo{pages}{071903} (\bibinfo{year}{2019}).

\bibitem[{\citenamefont{Jiang et~al.}(2021)\citenamefont{Jiang, Wen, Wang, Yin,
  Cheng, Liu, Feng, and He}}]{Jiang2021}
\bibinfo{author}{\bibfnamefont{J.}~\bibnamefont{Jiang}},
  \bibinfo{author}{\bibfnamefont{Y.}~\bibnamefont{Wen}},
  \bibinfo{author}{\bibfnamefont{H.}~\bibnamefont{Wang}},
  \bibinfo{author}{\bibfnamefont{L.}~\bibnamefont{Yin}},
  \bibinfo{author}{\bibfnamefont{R.}~\bibnamefont{Cheng}},
  \bibinfo{author}{\bibfnamefont{C.}~\bibnamefont{Liu}},
  \bibinfo{author}{\bibfnamefont{L.}~\bibnamefont{Feng}}, \bibnamefont{and}
  \bibinfo{author}{\bibfnamefont{J.}~\bibnamefont{He}},
  \bibinfo{journal}{Advanced Electronic Materials}
  \textbf{\bibinfo{volume}{7}}, \bibinfo{pages}{2001125}
  (\bibinfo{year}{2021}).

\bibitem[{\citenamefont{Ponc{\'{e}} et~al.}(2020)\citenamefont{Ponc{\'{e}}, Li,
  Reichardt, and Giustino}}]{Ponce2020}
\bibinfo{author}{\bibfnamefont{S.}~\bibnamefont{Ponc{\'{e}}}},
  \bibinfo{author}{\bibfnamefont{W.}~\bibnamefont{Li}},
  \bibinfo{author}{\bibfnamefont{S.}~\bibnamefont{Reichardt}},
  \bibnamefont{and} \bibinfo{author}{\bibfnamefont{F.}~\bibnamefont{Giustino}},
  \bibinfo{journal}{Reports on Progress in Physics}
  \textbf{\bibinfo{volume}{83}}, \bibinfo{pages}{036501}
  (\bibinfo{year}{2020}).

\bibitem[{\citenamefont{Ponc\'e et~al.}(2018)\citenamefont{Ponc\'e, Margine,
  and Giustino}}]{Ponce2018}
\bibinfo{author}{\bibfnamefont{S.}~\bibnamefont{Ponc\'e}},
  \bibinfo{author}{\bibfnamefont{E.~R.} \bibnamefont{Margine}},
  \bibnamefont{and} \bibinfo{author}{\bibfnamefont{F.}~\bibnamefont{Giustino}},
  \bibinfo{journal}{Phys. Rev. B} \textbf{\bibinfo{volume}{97}},
  \bibinfo{pages}{121201} (\bibinfo{year}{2018}).

\bibitem[{\citenamefont{Gonze and Lee}(1997)}]{Gonze1997a}
\bibinfo{author}{\bibfnamefont{X.}~\bibnamefont{Gonze}} \bibnamefont{and}
  \bibinfo{author}{\bibfnamefont{C.}~\bibnamefont{Lee}},
  \bibinfo{journal}{Phys. Rev. B} \textbf{\bibinfo{volume}{55}},
  \bibinfo{pages}{10355} (\bibinfo{year}{1997}).

\bibitem[{\citenamefont{Baroni et~al.}(2001)\citenamefont{Baroni, de~Gironcoli,
  Dal~Corso, and Giannozzi}}]{Baroni2001}
\bibinfo{author}{\bibfnamefont{S.}~\bibnamefont{Baroni}},
  \bibinfo{author}{\bibfnamefont{S.}~\bibnamefont{de~Gironcoli}},
  \bibinfo{author}{\bibfnamefont{A.}~\bibnamefont{Dal~Corso}},
  \bibnamefont{and}
  \bibinfo{author}{\bibfnamefont{P.}~\bibnamefont{Giannozzi}},
  \bibinfo{journal}{Rev. Mod. Phys.} \textbf{\bibinfo{volume}{73}},
  \bibinfo{pages}{515} (\bibinfo{year}{2001}).

\bibitem[{\citenamefont{Giustino et~al.}(2007)\citenamefont{Giustino, Cohen,
  and Louie}}]{Giustino2007}
\bibinfo{author}{\bibfnamefont{F.}~\bibnamefont{Giustino}},
  \bibinfo{author}{\bibfnamefont{M.~L.} \bibnamefont{Cohen}}, \bibnamefont{and}
  \bibinfo{author}{\bibfnamefont{S.~G.} \bibnamefont{Louie}},
  \bibinfo{journal}{Phys. Rev. B} \textbf{\bibinfo{volume}{76}},
  \bibinfo{pages}{165108} (\bibinfo{year}{2007}).

\bibitem[{\citenamefont{Eiguren and Ambrosch-Draxl}(2008)}]{Eiguren2008}
\bibinfo{author}{\bibfnamefont{A.}~\bibnamefont{Eiguren}} \bibnamefont{and}
  \bibinfo{author}{\bibfnamefont{C.}~\bibnamefont{Ambrosch-Draxl}},
  \bibinfo{journal}{Phys. Rev. B} \textbf{\bibinfo{volume}{78}},
  \bibinfo{pages}{045124} (\bibinfo{year}{2008}),
  \urlprefix\url{http://link.aps.org/doi/10.1103/PhysRevB.78.045124}.

\bibitem[{\citenamefont{Calandra et~al.}(2010)\citenamefont{Calandra, Profeta,
  and Mauri}}]{Calandra2010}
\bibinfo{author}{\bibfnamefont{M.}~\bibnamefont{Calandra}},
  \bibinfo{author}{\bibfnamefont{G.}~\bibnamefont{Profeta}}, \bibnamefont{and}
  \bibinfo{author}{\bibfnamefont{F.}~\bibnamefont{Mauri}},
  \bibinfo{journal}{Phys. Rev. B} \textbf{\bibinfo{volume}{82}},
  \bibinfo{pages}{165111} (\bibinfo{year}{2010}).

\bibitem[{\citenamefont{Brunin et~al.}(2020{\natexlab{a}})\citenamefont{Brunin,
  Miranda, Giantomassi, Royo, Stengel, Verstraete, Gonze, Rignanese, and
  Hautier}}]{Brunin2020}
\bibinfo{author}{\bibfnamefont{G.}~\bibnamefont{Brunin}},
  \bibinfo{author}{\bibfnamefont{H.~P.~C.} \bibnamefont{Miranda}},
  \bibinfo{author}{\bibfnamefont{M.}~\bibnamefont{Giantomassi}},
  \bibinfo{author}{\bibfnamefont{M.}~\bibnamefont{Royo}},
  \bibinfo{author}{\bibfnamefont{M.}~\bibnamefont{Stengel}},
  \bibinfo{author}{\bibfnamefont{M.~J.} \bibnamefont{Verstraete}},
  \bibinfo{author}{\bibfnamefont{X.}~\bibnamefont{Gonze}},
  \bibinfo{author}{\bibfnamefont{G.-M.} \bibnamefont{Rignanese}},
  \bibnamefont{and} \bibinfo{author}{\bibfnamefont{G.}~\bibnamefont{Hautier}},
  \bibinfo{journal}{Phys. Rev. B} \textbf{\bibinfo{volume}{102}},
  \bibinfo{pages}{094308} (\bibinfo{year}{2020}{\natexlab{a}}).

\bibitem[{\citenamefont{Vogl}(1976)}]{Vogl1976}
\bibinfo{author}{\bibfnamefont{P.}~\bibnamefont{Vogl}}, \bibinfo{journal}{Phys.
  Rev. B} \textbf{\bibinfo{volume}{13}}, \bibinfo{pages}{694}
  (\bibinfo{year}{1976}).

\bibitem[{\citenamefont{Brunin et~al.}(2020{\natexlab{b}})\citenamefont{Brunin,
  Miranda, Giantomassi, Royo, Stengel, Verstraete, Gonze, Rignanese, and
  Hautier}}]{Brunin2020a}
\bibinfo{author}{\bibfnamefont{G.}~\bibnamefont{Brunin}},
  \bibinfo{author}{\bibfnamefont{H.~P.~C.} \bibnamefont{Miranda}},
  \bibinfo{author}{\bibfnamefont{M.}~\bibnamefont{Giantomassi}},
  \bibinfo{author}{\bibfnamefont{M.}~\bibnamefont{Royo}},
  \bibinfo{author}{\bibfnamefont{M.}~\bibnamefont{Stengel}},
  \bibinfo{author}{\bibfnamefont{M.~J.} \bibnamefont{Verstraete}},
  \bibinfo{author}{\bibfnamefont{X.}~\bibnamefont{Gonze}},
  \bibinfo{author}{\bibfnamefont{G.-M.} \bibnamefont{Rignanese}},
  \bibnamefont{and} \bibinfo{author}{\bibfnamefont{G.}~\bibnamefont{Hautier}},
  \bibinfo{journal}{Phys. Rev. Lett.} \textbf{\bibinfo{volume}{125}},
  \bibinfo{pages}{136601} (\bibinfo{year}{2020}{\natexlab{b}}).

\bibitem[{\citenamefont{Jhalani et~al.}(2020)\citenamefont{Jhalani, Zhou, Park,
  Dreyer, and Bernardi}}]{Jhalani2020}
\bibinfo{author}{\bibfnamefont{V.~A.} \bibnamefont{Jhalani}},
  \bibinfo{author}{\bibfnamefont{J.-J.} \bibnamefont{Zhou}},
  \bibinfo{author}{\bibfnamefont{J.}~\bibnamefont{Park}},
  \bibinfo{author}{\bibfnamefont{C.~E.} \bibnamefont{Dreyer}},
  \bibnamefont{and} \bibinfo{author}{\bibfnamefont{M.}~\bibnamefont{Bernardi}},
  \bibinfo{journal}{Phys. Rev. Lett.} \textbf{\bibinfo{volume}{125}},
  \bibinfo{pages}{136602} (\bibinfo{year}{2020}).

\bibitem[{\citenamefont{Park et~al.}(2020)\citenamefont{Park, Zhou, Jhalani,
  Dreyer, and Bernardi}}]{Park2020a}
\bibinfo{author}{\bibfnamefont{J.}~\bibnamefont{Park}},
  \bibinfo{author}{\bibfnamefont{J.-J.} \bibnamefont{Zhou}},
  \bibinfo{author}{\bibfnamefont{V.~A.} \bibnamefont{Jhalani}},
  \bibinfo{author}{\bibfnamefont{C.~E.} \bibnamefont{Dreyer}},
  \bibnamefont{and} \bibinfo{author}{\bibfnamefont{M.}~\bibnamefont{Bernardi}},
  \bibinfo{journal}{Phys. Rev. B} \textbf{\bibinfo{volume}{102}},
  \bibinfo{pages}{125203} (\bibinfo{year}{2020}).

\bibitem[{\citenamefont{Ponc\'e et~al.}(2021)\citenamefont{Ponc\'e, Macheda,
  Margine, Marzari, Bonini, and Giustino}}]{Ponce2021}
\bibinfo{author}{\bibfnamefont{S.}~\bibnamefont{Ponc\'e}},
  \bibinfo{author}{\bibfnamefont{F.}~\bibnamefont{Macheda}},
  \bibinfo{author}{\bibfnamefont{E.~R.} \bibnamefont{Margine}},
  \bibinfo{author}{\bibfnamefont{N.}~\bibnamefont{Marzari}},
  \bibinfo{author}{\bibfnamefont{N.}~\bibnamefont{Bonini}}, \bibnamefont{and}
  \bibinfo{author}{\bibfnamefont{F.}~\bibnamefont{Giustino}},
  \bibinfo{journal}{Phys. Rev. Research} \textbf{\bibinfo{volume}{3}},
  \bibinfo{pages}{043022} (\bibinfo{year}{2021}).

\bibitem[{\citenamefont{Verdi and Giustino}(2015)}]{Verdi2015}
\bibinfo{author}{\bibfnamefont{C.}~\bibnamefont{Verdi}} \bibnamefont{and}
  \bibinfo{author}{\bibfnamefont{F.}~\bibnamefont{Giustino}},
  \bibinfo{journal}{Phys. Rev. Lett.} \textbf{\bibinfo{volume}{115}},
  \bibinfo{pages}{176401} (\bibinfo{year}{2015}).

\bibitem[{\citenamefont{Sjakste et~al.}(2015)\citenamefont{Sjakste, Vast,
  Calandra, and Mauri}}]{Sjakste2015}
\bibinfo{author}{\bibfnamefont{J.}~\bibnamefont{Sjakste}},
  \bibinfo{author}{\bibfnamefont{N.}~\bibnamefont{Vast}},
  \bibinfo{author}{\bibfnamefont{M.}~\bibnamefont{Calandra}}, \bibnamefont{and}
  \bibinfo{author}{\bibfnamefont{F.}~\bibnamefont{Mauri}},
  \bibinfo{journal}{Phys. Rev. B} \textbf{\bibinfo{volume}{92}},
  \bibinfo{pages}{054307} (\bibinfo{year}{2015}).

\bibitem[{\citenamefont{Giustino}(2017)}]{Giustino2017}
\bibinfo{author}{\bibfnamefont{F.}~\bibnamefont{Giustino}},
  \bibinfo{journal}{Rev. Mod. Phys.} \textbf{\bibinfo{volume}{89}},
  \bibinfo{pages}{015003} (\bibinfo{year}{2017}).

\bibitem[{\citenamefont{Sohier et~al.}(2016)\citenamefont{Sohier, Calandra, and
  Mauri}}]{Sohier2016}
\bibinfo{author}{\bibfnamefont{T.}~\bibnamefont{Sohier}},
  \bibinfo{author}{\bibfnamefont{M.}~\bibnamefont{Calandra}}, \bibnamefont{and}
  \bibinfo{author}{\bibfnamefont{F.}~\bibnamefont{Mauri}},
  \bibinfo{journal}{Phys. Rev. B} \textbf{\bibinfo{volume}{94}},
  \bibinfo{pages}{085415} (\bibinfo{year}{2016}).

\bibitem[{\citenamefont{Sohier et~al.}(2017{\natexlab{a}})\citenamefont{Sohier,
  Calandra, and Mauri}}]{Sohier2017}
\bibinfo{author}{\bibfnamefont{T.}~\bibnamefont{Sohier}},
  \bibinfo{author}{\bibfnamefont{M.}~\bibnamefont{Calandra}}, \bibnamefont{and}
  \bibinfo{author}{\bibfnamefont{F.}~\bibnamefont{Mauri}},
  \bibinfo{journal}{Phys. Rev. B} \textbf{\bibinfo{volume}{96}},
  \bibinfo{pages}{075448} (\bibinfo{year}{2017}{\natexlab{a}}).

\bibitem[{\citenamefont{Sohier et~al.}(2017{\natexlab{b}})\citenamefont{Sohier,
  Gibertini, Calandra, Mauri, and Marzari}}]{Sohier2017a}
\bibinfo{author}{\bibfnamefont{T.}~\bibnamefont{Sohier}},
  \bibinfo{author}{\bibfnamefont{M.}~\bibnamefont{Gibertini}},
  \bibinfo{author}{\bibfnamefont{M.}~\bibnamefont{Calandra}},
  \bibinfo{author}{\bibfnamefont{F.}~\bibnamefont{Mauri}}, \bibnamefont{and}
  \bibinfo{author}{\bibfnamefont{N.}~\bibnamefont{Marzari}},
  \bibinfo{journal}{Nano Letters} \textbf{\bibinfo{volume}{17}},
  \bibinfo{pages}{3758} (\bibinfo{year}{2017}{\natexlab{b}}).

\bibitem[{\citenamefont{Deng et~al.}(2021)\citenamefont{Deng, Wu, Shi, Wong,
  Wang, and Yang}}]{Deng2021}
\bibinfo{author}{\bibfnamefont{T.}~\bibnamefont{Deng}},
  \bibinfo{author}{\bibfnamefont{G.}~\bibnamefont{Wu}},
  \bibinfo{author}{\bibfnamefont{W.}~\bibnamefont{Shi}},
  \bibinfo{author}{\bibfnamefont{Z.~M.} \bibnamefont{Wong}},
  \bibinfo{author}{\bibfnamefont{J.-S.} \bibnamefont{Wang}}, \bibnamefont{and}
  \bibinfo{author}{\bibfnamefont{S.-W.} \bibnamefont{Yang}},
  \bibinfo{journal}{Phys. Rev. B} \textbf{\bibinfo{volume}{103}},
  \bibinfo{pages}{075410} (\bibinfo{year}{2021}).

\bibitem[{\citenamefont{Zhang and Liu}(2022)}]{Zhang2022}
\bibinfo{author}{\bibfnamefont{C.}~\bibnamefont{Zhang}} \bibnamefont{and}
  \bibinfo{author}{\bibfnamefont{Y.}~\bibnamefont{Liu}},
  \bibinfo{journal}{Phys. Rev. B} \textbf{\bibinfo{volume}{106}},
  \bibinfo{pages}{115423} (\bibinfo{year}{2022}),
  \urlprefix\url{https://link.aps.org/doi/10.1103/PhysRevB.106.115423}.

\bibitem[{\citenamefont{Marzari et~al.}(2012)\citenamefont{Marzari, Mostofi,
  Yates, Souza, and Vanderbilt}}]{Marzari2012}
\bibinfo{author}{\bibfnamefont{N.}~\bibnamefont{Marzari}},
  \bibinfo{author}{\bibfnamefont{A.~A.} \bibnamefont{Mostofi}},
  \bibinfo{author}{\bibfnamefont{J.~R.} \bibnamefont{Yates}},
  \bibinfo{author}{\bibfnamefont{I.}~\bibnamefont{Souza}}, \bibnamefont{and}
  \bibinfo{author}{\bibfnamefont{D.}~\bibnamefont{Vanderbilt}},
  \bibinfo{journal}{Rev. Mod. Phys.} \textbf{\bibinfo{volume}{84}},
  \bibinfo{pages}{1419} (\bibinfo{year}{2012}).

\bibitem[{\citenamefont{Ponc\'e et~al.}(2022)\citenamefont{Ponc\'e, Royo,
  Stengel, Marzari, and Gibertini}}]{Ponce2022}
\bibinfo{author}{\bibfnamefont{S.}~\bibnamefont{Ponc\'e}},
  \bibinfo{author}{\bibfnamefont{M.}~\bibnamefont{Royo}},
  \bibinfo{author}{\bibfnamefont{M.}~\bibnamefont{Stengel}},
  \bibinfo{author}{\bibfnamefont{N.}~\bibnamefont{Marzari}}, \bibnamefont{and}
  \bibinfo{author}{\bibfnamefont{M.}~\bibnamefont{Gibertini}},
  \bibinfo{journal}{XX} \textbf{\bibinfo{volume}{XX}}, \bibinfo{pages}{XXX}
  (\bibinfo{year}{2022}).

\bibitem[{\citenamefont{Royo and Stengel}(2021)}]{Royo2021}
\bibinfo{author}{\bibfnamefont{M.}~\bibnamefont{Royo}} \bibnamefont{and}
  \bibinfo{author}{\bibfnamefont{M.}~\bibnamefont{Stengel}},
  \bibinfo{journal}{Phys. Rev. X} \textbf{\bibinfo{volume}{11}},
  \bibinfo{pages}{041027} (\bibinfo{year}{2021}).

\bibitem[{\citenamefont{Stengel}(2013)}]{Stengel2013}
\bibinfo{author}{\bibfnamefont{M.}~\bibnamefont{Stengel}},
  \bibinfo{journal}{Phys. Rev. B} \textbf{\bibinfo{volume}{88}},
  \bibinfo{pages}{174106} (\bibinfo{year}{2013}).

\bibitem[{\citenamefont{Royo and Stengel}(2019)}]{Royo2019}
\bibinfo{author}{\bibfnamefont{M.}~\bibnamefont{Royo}} \bibnamefont{and}
  \bibinfo{author}{\bibfnamefont{M.}~\bibnamefont{Stengel}},
  \bibinfo{journal}{Phys. Rev. X} \textbf{\bibinfo{volume}{9}},
  \bibinfo{pages}{021050} (\bibinfo{year}{2019}).

\bibitem[{\citenamefont{Yu et~al.}(2016)\citenamefont{Yu, Ong, Pan, Cui, Xin,
  Shi, Wang, Wu, Chen, Zhang et~al.}}]{Yu2016}
\bibinfo{author}{\bibfnamefont{Z.}~\bibnamefont{Yu}},
  \bibinfo{author}{\bibfnamefont{Z.-Y.} \bibnamefont{Ong}},
  \bibinfo{author}{\bibfnamefont{Y.}~\bibnamefont{Pan}},
  \bibinfo{author}{\bibfnamefont{Y.}~\bibnamefont{Cui}},
  \bibinfo{author}{\bibfnamefont{R.}~\bibnamefont{Xin}},
  \bibinfo{author}{\bibfnamefont{Y.}~\bibnamefont{Shi}},
  \bibinfo{author}{\bibfnamefont{B.}~\bibnamefont{Wang}},
  \bibinfo{author}{\bibfnamefont{Y.}~\bibnamefont{Wu}},
  \bibinfo{author}{\bibfnamefont{T.}~\bibnamefont{Chen}},
  \bibinfo{author}{\bibfnamefont{Y.-W.} \bibnamefont{Zhang}},
  \bibnamefont{et~al.}, \bibinfo{journal}{Advanced Materials}
  \textbf{\bibinfo{volume}{28}}, \bibinfo{pages}{547} (\bibinfo{year}{2016}).

\bibitem[{\citenamefont{Cui et~al.}(2015)\citenamefont{Cui, Lee, Kim, Arefe,
  Huang, Lee, Chenet, Zhang, Wang, Ye et~al.}}]{Cui2015}
\bibinfo{author}{\bibfnamefont{X.}~\bibnamefont{Cui}},
  \bibinfo{author}{\bibfnamefont{G.-H.} \bibnamefont{Lee}},
  \bibinfo{author}{\bibfnamefont{Y.~D.} \bibnamefont{Kim}},
  \bibinfo{author}{\bibfnamefont{G.}~\bibnamefont{Arefe}},
  \bibinfo{author}{\bibfnamefont{P.~Y.} \bibnamefont{Huang}},
  \bibinfo{author}{\bibfnamefont{C.-H.} \bibnamefont{Lee}},
  \bibinfo{author}{\bibfnamefont{D.~A.} \bibnamefont{Chenet}},
  \bibinfo{author}{\bibfnamefont{X.}~\bibnamefont{Zhang}},
  \bibinfo{author}{\bibfnamefont{L.}~\bibnamefont{Wang}},
  \bibinfo{author}{\bibfnamefont{F.}~\bibnamefont{Ye}}, \bibnamefont{et~al.},
  \bibinfo{journal}{Nature Nanotechnology} \textbf{\bibinfo{volume}{10}},
  \bibinfo{pages}{534} (\bibinfo{year}{2015}).

\bibitem[{\citenamefont{Liu et~al.}(2015)\citenamefont{Liu, Wu, Cheng, Yang,
  Zhu, He, Ding, Li, O., Huang et~al.}}]{Liu2015c}
\bibinfo{author}{\bibfnamefont{Y.}~\bibnamefont{Liu}},
  \bibinfo{author}{\bibfnamefont{H.}~\bibnamefont{Wu}},
  \bibinfo{author}{\bibfnamefont{H.-C.} \bibnamefont{Cheng}},
  \bibinfo{author}{\bibfnamefont{S.}~\bibnamefont{Yang}},
  \bibinfo{author}{\bibfnamefont{E.}~\bibnamefont{Zhu}},
  \bibinfo{author}{\bibfnamefont{Q.}~\bibnamefont{He}},
  \bibinfo{author}{\bibfnamefont{M.}~\bibnamefont{Ding}},
  \bibinfo{author}{\bibfnamefont{J.}~\bibnamefont{Li}, \bibfnamefont{D.~Guo}},
  \bibinfo{author}{\bibfnamefont{W.~N.} \bibnamefont{O.}},
  \bibinfo{author}{\bibfnamefont{Y.}~\bibnamefont{Huang}},
  \bibnamefont{et~al.}, \bibinfo{journal}{Nano Lett.}
  \textbf{\bibinfo{volume}{15}}, \bibinfo{pages}{3030} (\bibinfo{year}{2015}).

\bibitem[{\citenamefont{Momose et~al.}(2018)\citenamefont{Momose, Nakamura,
  Daniel, and Shimomura}}]{Momose2018}
\bibinfo{author}{\bibfnamefont{T.}~\bibnamefont{Momose}},
  \bibinfo{author}{\bibfnamefont{A.}~\bibnamefont{Nakamura}},
  \bibinfo{author}{\bibfnamefont{M.}~\bibnamefont{Daniel}}, \bibnamefont{and}
  \bibinfo{author}{\bibfnamefont{M.}~\bibnamefont{Shimomura}},
  \bibinfo{journal}{AIP Advances} \textbf{\bibinfo{volume}{8}},
  \bibinfo{pages}{025009} (\bibinfo{year}{2018}).

\bibitem[{\citenamefont{Macheda et~al.}(2020)\citenamefont{Macheda, Ponc\'e,
  Giustino, and Bonini}}]{Macheda2020}
\bibinfo{author}{\bibfnamefont{F.}~\bibnamefont{Macheda}},
  \bibinfo{author}{\bibfnamefont{S.}~\bibnamefont{Ponc\'e}},
  \bibinfo{author}{\bibfnamefont{F.}~\bibnamefont{Giustino}}, \bibnamefont{and}
  \bibinfo{author}{\bibfnamefont{N.}~\bibnamefont{Bonini}},
  \bibinfo{journal}{Nano Letters} \textbf{\bibinfo{volume}{20}},
  \bibinfo{pages}{8861} (\bibinfo{year}{2020}).

\bibitem[{\citenamefont{Reggiani et~al.}(1983)\citenamefont{Reggiani, Waechter,
  and Zukotynski}}]{Reggiani1983}
\bibinfo{author}{\bibfnamefont{L.}~\bibnamefont{Reggiani}},
  \bibinfo{author}{\bibfnamefont{D.}~\bibnamefont{Waechter}}, \bibnamefont{and}
  \bibinfo{author}{\bibfnamefont{S.}~\bibnamefont{Zukotynski}},
  \bibinfo{journal}{Phys. Rev. B} \textbf{\bibinfo{volume}{28}},
  \bibinfo{pages}{3550} (\bibinfo{year}{1983}).

\bibitem[{\citenamefont{Popovic}(1991)}]{Popovic1991}
\bibinfo{author}{\bibfnamefont{R.}~\bibnamefont{Popovic}},
  \emph{\bibinfo{title}{Hall Effect Devices: Magnetic Sensors and
  Characterization of Semiconductors}} (\bibinfo{publisher}{Taylor \& Francis:
  Boca Raton}, \bibinfo{year}{1991}).

\bibitem[{\citenamefont{Macheda and Bonini}(2018)}]{Macheda2018}
\bibinfo{author}{\bibfnamefont{F.}~\bibnamefont{Macheda}} \bibnamefont{and}
  \bibinfo{author}{\bibfnamefont{N.}~\bibnamefont{Bonini}},
  \bibinfo{journal}{Phys. Rev. B} \textbf{\bibinfo{volume}{98}},
  \bibinfo{pages}{201201(R)} (\bibinfo{year}{2018}).

\bibitem[{\citenamefont{Ponc\'e et~al.}(2016)\citenamefont{Ponc\'e, Margine,
  Verdi, and Giustino}}]{Ponce2016a}
\bibinfo{author}{\bibfnamefont{S.}~\bibnamefont{Ponc\'e}},
  \bibinfo{author}{\bibfnamefont{E.}~\bibnamefont{Margine}},
  \bibinfo{author}{\bibfnamefont{C.}~\bibnamefont{Verdi}}, \bibnamefont{and}
  \bibinfo{author}{\bibfnamefont{F.}~\bibnamefont{Giustino}},
  \bibinfo{journal}{Computer Physics Communications}
  \textbf{\bibinfo{volume}{209}}, \bibinfo{pages}{116 } (\bibinfo{year}{2016}).

\bibitem[{\citenamefont{Pizzi et~al.}(2020)\citenamefont{Pizzi, Vitale, Arita,
  Bl{\"u}gel, Freimuth, G{\'{e}}ranton, Gibertini, Gresch, Johnson, Koretsune
  et~al.}}]{Pizzi2020}
\bibinfo{author}{\bibfnamefont{G.}~\bibnamefont{Pizzi}},
  \bibinfo{author}{\bibfnamefont{V.}~\bibnamefont{Vitale}},
  \bibinfo{author}{\bibfnamefont{R.}~\bibnamefont{Arita}},
  \bibinfo{author}{\bibfnamefont{S.}~\bibnamefont{Bl{\"u}gel}},
  \bibinfo{author}{\bibfnamefont{F.}~\bibnamefont{Freimuth}},
  \bibinfo{author}{\bibfnamefont{G.}~\bibnamefont{G{\'{e}}ranton}},
  \bibinfo{author}{\bibfnamefont{M.}~\bibnamefont{Gibertini}},
  \bibinfo{author}{\bibfnamefont{D.}~\bibnamefont{Gresch}},
  \bibinfo{author}{\bibfnamefont{C.}~\bibnamefont{Johnson}},
  \bibinfo{author}{\bibfnamefont{T.}~\bibnamefont{Koretsune}},
  \bibnamefont{et~al.}, \bibinfo{journal}{Journal of Physics: Condensed Matter}
  \textbf{\bibinfo{volume}{32}}, \bibinfo{pages}{165902}
  (\bibinfo{year}{2020}).

\bibitem[{\citenamefont{Giannozzi et~al.}(2017)\citenamefont{Giannozzi,
  Andreussi, Brumme, Bunau, Nardelli, Calandra, Car, Cavazzoni, Ceresoli,
  Cococcioni et~al.}}]{Giannozzi2017}
\bibinfo{author}{\bibfnamefont{P.}~\bibnamefont{Giannozzi}},
  \bibinfo{author}{\bibfnamefont{O.}~\bibnamefont{Andreussi}},
  \bibinfo{author}{\bibfnamefont{T.}~\bibnamefont{Brumme}},
  \bibinfo{author}{\bibfnamefont{O.}~\bibnamefont{Bunau}},
  \bibinfo{author}{\bibfnamefont{M.~B.} \bibnamefont{Nardelli}},
  \bibinfo{author}{\bibfnamefont{M.}~\bibnamefont{Calandra}},
  \bibinfo{author}{\bibfnamefont{R.}~\bibnamefont{Car}},
  \bibinfo{author}{\bibfnamefont{C.}~\bibnamefont{Cavazzoni}},
  \bibinfo{author}{\bibfnamefont{D.}~\bibnamefont{Ceresoli}},
  \bibinfo{author}{\bibfnamefont{M.}~\bibnamefont{Cococcioni}},
  \bibnamefont{et~al.}, \bibinfo{journal}{Journal of Physics: Condensed Matter}
  \textbf{\bibinfo{volume}{29}}, \bibinfo{pages}{465901}
  (\bibinfo{year}{2017}).

\bibitem[{\citenamefont{Gonze et~al.}(2016)\citenamefont{Gonze, Jollet, Araujo,
  Adams, Amadon, Applencourt, Audouze, Beuken, Bieder, Bokhanchuk
  et~al.}}]{Gonze2016}
\bibinfo{author}{\bibfnamefont{X.}~\bibnamefont{Gonze}},
  \bibinfo{author}{\bibfnamefont{F.}~\bibnamefont{Jollet}},
  \bibinfo{author}{\bibfnamefont{F.~A.} \bibnamefont{Araujo}},
  \bibinfo{author}{\bibfnamefont{D.}~\bibnamefont{Adams}},
  \bibinfo{author}{\bibfnamefont{B.}~\bibnamefont{Amadon}},
  \bibinfo{author}{\bibfnamefont{T.}~\bibnamefont{Applencourt}},
  \bibinfo{author}{\bibfnamefont{C.}~\bibnamefont{Audouze}},
  \bibinfo{author}{\bibfnamefont{J.-M.} \bibnamefont{Beuken}},
  \bibinfo{author}{\bibfnamefont{J.}~\bibnamefont{Bieder}},
  \bibinfo{author}{\bibfnamefont{A.}~\bibnamefont{Bokhanchuk}},
  \bibnamefont{et~al.}, \bibinfo{journal}{Computer Physics Communications}
  \textbf{\bibinfo{volume}{205}}, \bibinfo{pages}{106 } (\bibinfo{year}{2016}).

\bibitem[{\citenamefont{Gonze et~al.}(2020)\citenamefont{Gonze, Amadon,
  Antonius, Arnardi, Baguet, Beuken, Bieder, Bottin, Bouchet, Bousquet
  et~al.}}]{Gonze2020}
\bibinfo{author}{\bibfnamefont{X.}~\bibnamefont{Gonze}},
  \bibinfo{author}{\bibfnamefont{B.}~\bibnamefont{Amadon}},
  \bibinfo{author}{\bibfnamefont{G.}~\bibnamefont{Antonius}},
  \bibinfo{author}{\bibfnamefont{F.}~\bibnamefont{Arnardi}},
  \bibinfo{author}{\bibfnamefont{L.}~\bibnamefont{Baguet}},
  \bibinfo{author}{\bibfnamefont{J.-M.} \bibnamefont{Beuken}},
  \bibinfo{author}{\bibfnamefont{J.}~\bibnamefont{Bieder}},
  \bibinfo{author}{\bibfnamefont{F.}~\bibnamefont{Bottin}},
  \bibinfo{author}{\bibfnamefont{J.}~\bibnamefont{Bouchet}},
  \bibinfo{author}{\bibfnamefont{E.}~\bibnamefont{Bousquet}},
  \bibnamefont{et~al.}, \bibinfo{journal}{Comput. Phys. Commun.}
  \textbf{\bibinfo{volume}{248}}, \bibinfo{pages}{107042}
  (\bibinfo{year}{2020}).

\bibitem[{\citenamefont{Zollner et~al.}(1990)\citenamefont{Zollner, Gopalan,
  and Cardona}}]{Zollner1990}
\bibinfo{author}{\bibfnamefont{S.}~\bibnamefont{Zollner}},
  \bibinfo{author}{\bibfnamefont{S.}~\bibnamefont{Gopalan}}, \bibnamefont{and}
  \bibinfo{author}{\bibfnamefont{M.}~\bibnamefont{Cardona}},
  \bibinfo{journal}{Journal of Applied Physics} \textbf{\bibinfo{volume}{68}},
  \bibinfo{pages}{1682} (\bibinfo{year}{1990}).

\bibitem[{\citenamefont{Yu et~al.}(2014)\citenamefont{Yu, Pan, Shen, Wang, Ong,
  Xu, Xin, Pan, Wang, Sun et~al.}}]{Yu2014}
\bibinfo{author}{\bibfnamefont{Z.}~\bibnamefont{Yu}},
  \bibinfo{author}{\bibfnamefont{Y.}~\bibnamefont{Pan}},
  \bibinfo{author}{\bibfnamefont{Y.}~\bibnamefont{Shen}},
  \bibinfo{author}{\bibfnamefont{Z.}~\bibnamefont{Wang}},
  \bibinfo{author}{\bibfnamefont{Z.~Y.} \bibnamefont{Ong}},
  \bibinfo{author}{\bibfnamefont{T.}~\bibnamefont{Xu}},
  \bibinfo{author}{\bibfnamefont{R.}~\bibnamefont{Xin}},
  \bibinfo{author}{\bibfnamefont{L.}~\bibnamefont{Pan}},
  \bibinfo{author}{\bibfnamefont{B.}~\bibnamefont{Wang}},
  \bibinfo{author}{\bibfnamefont{L.}~\bibnamefont{Sun}}, \bibnamefont{et~al.},
  \bibinfo{journal}{Nature Communications} \textbf{\bibinfo{volume}{5}},
  \bibinfo{pages}{5290} (\bibinfo{year}{2014}).

\bibitem[{\citenamefont{Sanne et~al.}(2015)\citenamefont{Sanne, Ghosh, Rai,
  Movva, Sharma, Rao, Mathew, and Banerjee}}]{Sanne2015}
\bibinfo{author}{\bibfnamefont{A.}~\bibnamefont{Sanne}},
  \bibinfo{author}{\bibfnamefont{R.}~\bibnamefont{Ghosh}},
  \bibinfo{author}{\bibfnamefont{A.}~\bibnamefont{Rai}},
  \bibinfo{author}{\bibfnamefont{H.~C.~P.} \bibnamefont{Movva}},
  \bibinfo{author}{\bibfnamefont{A.}~\bibnamefont{Sharma}},
  \bibinfo{author}{\bibfnamefont{R.}~\bibnamefont{Rao}},
  \bibinfo{author}{\bibfnamefont{L.}~\bibnamefont{Mathew}}, \bibnamefont{and}
  \bibinfo{author}{\bibfnamefont{S.~K.} \bibnamefont{Banerjee}},
  \bibinfo{journal}{Applied Physics Letters} \textbf{\bibinfo{volume}{106}},
  \bibinfo{pages}{062101} (\bibinfo{year}{2015}).

\bibitem[{\citenamefont{Kang et~al.}(2015)\citenamefont{Kang, Xie, Huang, Han,
  Huang, Mak, Kim, Muller, and Park}}]{Kang2015}
\bibinfo{author}{\bibfnamefont{K.}~\bibnamefont{Kang}},
  \bibinfo{author}{\bibfnamefont{S.}~\bibnamefont{Xie}},
  \bibinfo{author}{\bibfnamefont{L.}~\bibnamefont{Huang}},
  \bibinfo{author}{\bibfnamefont{Y.}~\bibnamefont{Han}},
  \bibinfo{author}{\bibfnamefont{P.~Y.} \bibnamefont{Huang}},
  \bibinfo{author}{\bibfnamefont{K.~F.} \bibnamefont{Mak}},
  \bibinfo{author}{\bibfnamefont{C.~J.} \bibnamefont{Kim}},
  \bibinfo{author}{\bibfnamefont{D.}~\bibnamefont{Muller}}, \bibnamefont{and}
  \bibinfo{author}{\bibfnamefont{J.}~\bibnamefont{Park}},
  \bibinfo{journal}{Nature} \textbf{\bibinfo{volume}{520}},
  \bibinfo{pages}{656} (\bibinfo{year}{2015}).

\bibitem[{\citenamefont{Kaasbjerg et~al.}(2012)\citenamefont{Kaasbjerg,
  Thygesen, and Jacobsen}}]{Kaasbjerg2012a}
\bibinfo{author}{\bibfnamefont{K.}~\bibnamefont{Kaasbjerg}},
  \bibinfo{author}{\bibfnamefont{K.~S.} \bibnamefont{Thygesen}},
  \bibnamefont{and} \bibinfo{author}{\bibfnamefont{K.~W.}
  \bibnamefont{Jacobsen}}, \bibinfo{journal}{Phys. Rev. B}
  \textbf{\bibinfo{volume}{85}}, \bibinfo{pages}{115317}
  (\bibinfo{year}{2012}).

\bibitem[{\citenamefont{Li et~al.}(2013)\citenamefont{Li, Mullen, Jin,
  Borysenko, Buongiorno~Nardelli, and Kim}}]{Li2013}
\bibinfo{author}{\bibfnamefont{X.}~\bibnamefont{Li}},
  \bibinfo{author}{\bibfnamefont{J.~T.} \bibnamefont{Mullen}},
  \bibinfo{author}{\bibfnamefont{Z.}~\bibnamefont{Jin}},
  \bibinfo{author}{\bibfnamefont{K.~M.} \bibnamefont{Borysenko}},
  \bibinfo{author}{\bibfnamefont{M.}~\bibnamefont{Buongiorno~Nardelli}},
  \bibnamefont{and} \bibinfo{author}{\bibfnamefont{K.~W.} \bibnamefont{Kim}},
  \bibinfo{journal}{Phys. Rev. B} \textbf{\bibinfo{volume}{87}},
  \bibinfo{pages}{115418} (\bibinfo{year}{2013}).

\bibitem[{\citenamefont{Kaasbjerg et~al.}(2013)\citenamefont{Kaasbjerg,
  Thygesen, and Jauho}}]{Kaasbjerg2013}
\bibinfo{author}{\bibfnamefont{K.}~\bibnamefont{Kaasbjerg}},
  \bibinfo{author}{\bibfnamefont{K.~S.} \bibnamefont{Thygesen}},
  \bibnamefont{and} \bibinfo{author}{\bibfnamefont{A.-P.} \bibnamefont{Jauho}},
  \bibinfo{journal}{Phys. Rev. B} \textbf{\bibinfo{volume}{87}},
  \bibinfo{pages}{235312} (\bibinfo{year}{2013}).

\bibitem[{\citenamefont{Zhang et~al.}(2014)\citenamefont{Zhang, Huang, Zhang,
  and Li}}]{Zhang2014}
\bibinfo{author}{\bibfnamefont{W.}~\bibnamefont{Zhang}},
  \bibinfo{author}{\bibfnamefont{Z.}~\bibnamefont{Huang}},
  \bibinfo{author}{\bibfnamefont{W.}~\bibnamefont{Zhang}}, \bibnamefont{and}
  \bibinfo{author}{\bibfnamefont{Y.}~\bibnamefont{Li}}, \bibinfo{journal}{Nano
  Research} \textbf{\bibinfo{volume}{7}}, \bibinfo{pages}{1731}
  (\bibinfo{year}{2014}).

\bibitem[{\citenamefont{Li}(2015)}]{Li2015}
\bibinfo{author}{\bibfnamefont{W.}~\bibnamefont{Li}}, \bibinfo{journal}{Phys.
  Rev. B} \textbf{\bibinfo{volume}{92}}, \bibinfo{pages}{075405}
  (\bibinfo{year}{2015}).

\bibitem[{\citenamefont{Gunst et~al.}(2016)\citenamefont{Gunst, Markussen,
  Stokbro, and Brandbyge}}]{Gunst2016}
\bibinfo{author}{\bibfnamefont{T.}~\bibnamefont{Gunst}},
  \bibinfo{author}{\bibfnamefont{T.}~\bibnamefont{Markussen}},
  \bibinfo{author}{\bibfnamefont{K.}~\bibnamefont{Stokbro}}, \bibnamefont{and}
  \bibinfo{author}{\bibfnamefont{M.}~\bibnamefont{Brandbyge}},
  \bibinfo{journal}{Phys. Rev. B} \textbf{\bibinfo{volume}{93}},
  \bibinfo{pages}{035414} (\bibinfo{year}{2016}).

\bibitem[{\citenamefont{Sohier et~al.}(2018)\citenamefont{Sohier, Campi,
  Marzari, and Gibertini}}]{Sohier2018}
\bibinfo{author}{\bibfnamefont{T.}~\bibnamefont{Sohier}},
  \bibinfo{author}{\bibfnamefont{D.}~\bibnamefont{Campi}},
  \bibinfo{author}{\bibfnamefont{N.}~\bibnamefont{Marzari}}, \bibnamefont{and}
  \bibinfo{author}{\bibfnamefont{M.}~\bibnamefont{Gibertini}},
  \bibinfo{journal}{Phys. Rev. Materials} \textbf{\bibinfo{volume}{2}},
  \bibinfo{pages}{114010} (\bibinfo{year}{2018}).

\bibitem[{\citenamefont{Guo et~al.}(2019)\citenamefont{Guo, Liu, Zhu, and
  Zheng}}]{Guo2019}
\bibinfo{author}{\bibfnamefont{F.}~\bibnamefont{Guo}},
  \bibinfo{author}{\bibfnamefont{Z.}~\bibnamefont{Liu}},
  \bibinfo{author}{\bibfnamefont{M.}~\bibnamefont{Zhu}}, \bibnamefont{and}
  \bibinfo{author}{\bibfnamefont{Y.}~\bibnamefont{Zheng}},
  \bibinfo{journal}{Phys. Chem. Chem. Phys.} \textbf{\bibinfo{volume}{21}},
  \bibinfo{pages}{22879} (\bibinfo{year}{2019}).

\bibitem[{\citenamefont{Gaddemane et~al.}(2021)\citenamefont{Gaddemane,
  Gopalan, de~Put, and Fischetti}}]{Gaddemane2019}
\bibinfo{author}{\bibfnamefont{G.}~\bibnamefont{Gaddemane}},
  \bibinfo{author}{\bibfnamefont{S.}~\bibnamefont{Gopalan}},
  \bibinfo{author}{\bibfnamefont{M.~V.} \bibnamefont{de~Put}},
  \bibnamefont{and} \bibinfo{author}{\bibfnamefont{M.~V.}
  \bibnamefont{Fischetti}}, \bibinfo{journal}{J. Comput. Elect.}
  \textbf{\bibinfo{volume}{20}}, \bibinfo{pages}{49} (\bibinfo{year}{2021}).

\bibitem[{IRD()}]{IRDS2001}
\emph{\bibinfo{title}{International roadmap for devices and systems (irds)}}.

\bibitem[{\citenamefont{Perdew et~al.}(1996)\citenamefont{Perdew, Burke, and
  Ernzerhof}}]{Perdew1996}
\bibinfo{author}{\bibfnamefont{J.~P.} \bibnamefont{Perdew}},
  \bibinfo{author}{\bibfnamefont{K.}~\bibnamefont{Burke}}, \bibnamefont{and}
  \bibinfo{author}{\bibfnamefont{M.}~\bibnamefont{Ernzerhof}},
  \bibinfo{journal}{Phys. Rev. Lett.} \textbf{\bibinfo{volume}{77}},
  \bibinfo{pages}{3865} (\bibinfo{year}{1996}).

\bibitem[{\citenamefont{Hamann}(2013)}]{Hamann2013}
\bibinfo{author}{\bibfnamefont{D.~R.} \bibnamefont{Hamann}},
  \bibinfo{journal}{Phys. Rev. B} \textbf{\bibinfo{volume}{88}},
  \bibinfo{pages}{085117} (\bibinfo{year}{2013}).

\bibitem[{\citenamefont{van Setten et~al.}(2018)\citenamefont{van Setten,
  Giantomassi, Bousquet, Verstraete, Hamann, Gonze, and
  Rignanese}}]{Setten2018}
\bibinfo{author}{\bibfnamefont{M.}~\bibnamefont{van Setten}},
  \bibinfo{author}{\bibfnamefont{M.}~\bibnamefont{Giantomassi}},
  \bibinfo{author}{\bibfnamefont{E.}~\bibnamefont{Bousquet}},
  \bibinfo{author}{\bibfnamefont{M.}~\bibnamefont{Verstraete}},
  \bibinfo{author}{\bibfnamefont{D.}~\bibnamefont{Hamann}},
  \bibinfo{author}{\bibfnamefont{X.}~\bibnamefont{Gonze}}, \bibnamefont{and}
  \bibinfo{author}{\bibfnamefont{G.-M.} \bibnamefont{Rignanese}},
  \bibinfo{journal}{Computer Physics Communications}
  \textbf{\bibinfo{volume}{226}}, \bibinfo{pages}{39 } (\bibinfo{year}{2018}).

\end{thebibliography}
\end{document}